\documentclass[prb,twocolumn,groupedaddress]{revtex4}
\usepackage{graphicx}
\usepackage{graphics}
\usepackage{amssymb,amsmath}
\usepackage[utf8x]{inputenc}
\usepackage{dcolumn}% Align table columns on decimal point
\usepackage{bm}% bold math                
\usepackage{epsfig}
\usepackage{psfrag}
\usepackage{yfonts}
\usepackage{version}
\usepackage{natbib}

\newcommand{\fref}[1]{Fig.~\ref{#1}}

\newcommand{\eref}[1]{Eq.~(\ref{#1})}

\newcommand{\cref}[1]{chapter~\ref{#1}}

\newcommand{\Cref}[1]{Chapter~\ref{#1}}

\begin{document}

\title{Phase directed excitonic transport and its limitations due to
  environmental influence}

\author{ Alexander Eisfeld}
\affiliation{Max-Planck-Institut f\"ur Physik komplexer Systeme, N\"othnitzer Str. 38, D-01187 Dresden, Germany}

\begin{abstract}
We investigate theoretically the transfer of excitation along a one
dimensional chain of monomers for a situation in which initially the
excitation is shared coherently by two monomers.
We show that depending on the relative phase between the two
monomers strong directionality   of the energy transfer is possible.
It is also investigated how dephasing, induced by an environment, influences
this directed transport.   

\end{abstract}

\keywords{
  Frenkel exciton, energy transfer, J-aggregates, H-aggregates, dephasing}

%%%%%%%%%%%%%%%%%%%%%%%%%%%%%%%%%%%%%%%%%%%%%%%%%%%%%%%%%%%%%%%%%%%%%%%
%%%%%%%%%%%%%%%%%%%%%%%%%%%%%%%%%%%%%%%%%%%%%%%%%%%%%%%%%%%%%%%%%%%%%%%

\maketitle
%%%%%%%%%%%%%%%%%%%%%%%%%%%%%%%%%%%%%%%%%%%%%%%%%%%%%%%%%%%%%%%%%%%%%%5
\newcommand{\bra}[1]{\langle\,{#1}\, |}
\newcommand{\ket}[1]{|\,{#1}\,\rangle}
\newcommand{\braket}[2]{\mbox{$\langle\,{#1}\, | \,{#2}\,\rangle$}}
\newcommand{\ketbra}[2]{|\,{#1}\,\rangle\langle\,{#2}\,|}
\newcommand{\eps}{\mbox{\boldmath $\epsilon$}}

\newcommand{\lrb}[1]{\langle\, {#1}\,\rangle}
\newcommand{\Erw}[1]{\big[\!\!\big[{#1}\big]\!\!\big]}
\newcommand{\Real}{\mbox{Re}}

\newcommand{\V}{V}
\newcommand{\ElTransE}{\epsilon}%{\mathcal{E}}
\newcommand{\HamTot}{H}
\newcommand{\transEnergy}{\epsilon}
\newcommand{\meanTrans}{{\ElTransE}}
\newcommand{\width}{\sigma}
\newcommand{\variance}{\sigma^2}
\newcommand{\transitEAgg}{\ElTransE_{\rm \tiny Agg}}
\newcommand{\ScalePar}{\sigma}
\newcommand{\StabIndex}{\alpha}
\newcommand{\Distrib}{S}
\newcommand{\FWHM}{\Delta}
\newcommand{\p}{\mathcal{P}}
\newcommand{\W}{W}
\newcommand{\Kern}{Q}

\newcommand{\LichtPol} {\hat{\mathcal{E}}}  

\newcommand{ \dipOp} { \vec{D}}    % (dip)ole (o)perator
 
\newcommand{ \dip}{ \vec{d}}    
\newcommand{\e}{\mbox{e}}

\newcommand{\om}{\omega}

\newcommand{\al}{\alpha}
\newcommand{\cm}{\ {\rm cm}^{-1}}

\section{Introduction}
Controlling the direction of motion of a quantum mechanical wave packed has received much interest, e.g.\ in the context of unidirectional rotational motion\cite{HoYaKo03_12393_,MaGrSc06_054325_,PePeGo10_075007_}.
In the present work we will investigate unidirectional excitation transfer in assemblies of monomers that interact via resonant dipole-dipole interaction (see for instance Refs.~\cite{Me58_647_,Go63_1245_,GrSi71_4843_,Da62__,Ho66_208_,HaRe71_253_,HaSt73_135_,SiCa09_13825_,ReMoKa09_033003_,RoScEi09_044909_}).
Among the various examples  are molecular aggregates \cite{Ko96__}, organic molecular solids \cite{ScWo06__}, light harvesting systems \cite{AmVaGr00__}, chains of quantum dots, or  chains of ultra-cold Rydberg atoms \cite{WAtEi10_053004_,MueBlAm07_090601_,RoHeTo04_042703_}.
In all these systems the transport of electronic excitation can adequately be described
by the exciton theory of Frenkel \cite{Fr31_17_,Fr30_198_,FrTe38_861_}, where
electronic excitation localized on a monomer can be transferred to another
monomer due to transition dipole-dipole interaction.

In most of the studies of this excitation transfer usually the initial condition is
taken in such a way that the excitation is localized  on a {\it single} monomer,
i.e.\ one monomer is electronically excited and all the others
are in their electronic ground state (e.g.\ in Refs.~\cite{Me58_647_,RoScEi09_044909_,MaFu61_1715_,Bi67_1484_,RoEiWo09_058301_}) 
Very little attention is paid to  the investigation of initial states that
extend over several monomers. In this situation, initial phase relations between
different monomers play an important role \cite{Ja09_164101_}. 
With the increasing capability to create artificial structures, controlling the initial condition can be used to tailor the properties of such systems.
One possibility to create a state
  where excitation is distributed  coherently over several monomers, is to use
  electromagnetic radiation that has a definite phase relation between
  different monomers. Since for a typical  molecular aggregate\cite{Ko96__,EiKnKi09_658_,AmVaGr00__} the distance between the monomers is in the oder of a few {\AA}ngstr\"om and the excitation wavelength is in the visible, direct addressing of individual chromophores is quite difficult and would probably require the use of near-field methods.
% While direct addressing of individual chromophores in a typical   molecular aggregate\cite{Ko96__,EiKnKi09_658_,AmVaGr00__} will be a hard  task due to the small separation between the monomers, 
For chains of   ultra-cold Rydberg atoms, where the distances between the monomers can be in  the $\mu$m range \cite{UrJoHe09_110_,GaMiWi09_115_}, this addressing should  be easier to achiv and in principle be  possible with present-day technology.

In this work it is pointed out that,  even for an initial state in which the excitation is
delocalized only over two neighboring monomers, it is possible to obtain very
strong unidirectional transport along a one-dimensional chain by a suitable
choice of the relative phase between the excitations of neighboring monomers. 
To investigate the robustness of the directionality against fluctuations of
the surroundings
we describe the  interaction of the Frenkel exciton with its environment within the Haken-Reineker-Strobl model, including pure dephasing.

The paper is organized as follows:
In section \ref{sec:describ} we specify the system considered and 
the initial states.
In section \ref{sec:numerics} numerical calculations are performed to
illustrate the influence of the relative phase on the direction of transport.
Here also the interplay with the environment is investigated.
In section \ref{sec:analytics}  we give intuitive analytical explanations for the
observed directionality. In particular, an exact formula for the time-dependent mean-position of the excitation is derived, including
the interaction with the environment.
In section \ref{sec:conclusion} we conclude with a summary of our findings and give a brief outlook.

\section{Description of the system}
\label{sec:describ}
We consider a one dimensional array of $N$ monomers. The monomers are described
by electronic two-level systems, where $\ket{\phi_n^g}$ denotes the  ground
state and  $\ket{\phi_n^e}$  the excited state of monomer $n$. 
The energy difference between the two states is denoted by $\epsilon_n$. 
Since we focus on the transport of a single exctation, only aggregate stats of the form
\begin{equation}
  \label{eq:pi_n}
  \ket{\pi_n}=\ket{\phi_1^g}\cdots \ket{\phi_n^e}\cdots \ket{\phi_N^g},
\end{equation}
where monomer $n$ is excited and all the others are in their ground state, are taken into account.
%We consider  aggregate states where one excitation is in the system.
%A state in which monomer $n$ is excited and all the others are
%in their ground state is denoted by
%\begin{equation}
%  \label{eq:pi_n}
%  \ket{\pi_n}=\ket{\phi_1^g}\cdots \ket{\phi_n^e}\cdots \ket{\phi_N^g}
%\end{equation}
The aggregate is described by the Hamiltonian
%We  consider the following Hamiltonian corresponding to a simple  Frenkel exciton model \cite{MaKue00__,AmVaGr00__}
\begin{equation}
  \label{eq:Ham}
  H=\sum_{n=1}^N \epsilon_n \ket{\pi_n}\bra{\pi_n}+\sum_{n,m=1 \atop n\ne m}^N V_{nm}\ket{\pi_n}\bra{\pi_m}
\end{equation}
corresponding to a simple  Frenkel exciton model \cite{MaKue00__,AmVaGr00__}.
The transfer matrix element  $V_{nm}$ is typically taken to be of dipole-dipole form
with a distance dependence $V_{nm}\sim 1/|R_n-R_m|^3$, where $R_n$ and $R_m$ are the positions of monomer $n$ and $m$.
We focus on a one-dimensional regular chain of equidistantly spaced monomers and  use
\begin{equation}
  \label{eq:V_nm}
  V_{nm}=V \, \frac{1}{|n-m|^3}, \hspace{1cm} n\ne m
\end{equation}
where $V$ scales the interaction strength between the monomers.
In the following we take $\epsilon_n=\epsilon$ independent of the monomer
index $n$.

Following the approach of Haken, Reineker and Strobl
\cite{HaRe71_253_,HaSt73_135_},   
the interaction of the monomers with the surroundings is taken into account
by the  time dependent Hamiltonian
\begin{equation}
\label{eq:H'}
H'(t)=\sum_{n}q_n(t)\ket{\pi_n}\bra{\pi_n}
\end{equation}
Here $q_n(t)$ are stochastic environment fluctuations which couple to the excitation
on monomer $n$. They are taken to be real Gaussian random variables with 
mean
\begin{equation}
\label{eq:<qm>}
\langle q_n(t) \rangle=0
\end{equation}
and variance
\begin{equation}
\label{eq:<qnqm>}
\langle q_n(t) q_m(s)\rangle= \gamma_n \delta_{nm}\delta(t-s)
\end{equation}
The average over the fluctuations is denoted by  $\langle \cdots \rangle$. 
In Eq.~(\ref{eq:<qnqm>}) we have assumed that fluctuations at different
monomers $n$ and $m$ are uncorrelated and that the environment has no memory,
i.e.\ it is Markovian.
In recent years there has been much interest in the influence of non-Markovian
environments on energy transfer \cite{RoEiWo09_058301_,ReChAs09_184102_} but to discuss the basic influence of the environment on the directed transport, a
Markovian description will be sufficient. 
In the conclusions we will briefly discuss the effects of
  non-Markovian environments together with the influence of internal
  vibrations of the monomers. 
Furthermore we will assume that the  $\gamma_n$ are site independent,
i.e.\ $\gamma_n=\gamma$ for all $n$.

From the Liouville--von~Neumann equation,
averaged over the environmental fluctuations, one finds for the matrix
elements $\rho_{nm}=\bra{\pi_n}\rho\ket{\pi_m}$ of the averaged  density matrix \cite{HaRe71_253_}
\begin{equation}
  \label{eq:dot_rho_nm}
   \dot\rho_{nm}(t)=- \frac{i}{\hbar} [H,\rho(t)]_{nm}-\gamma\ (1- \delta_{nm})\rho_{nm}(t).  
\end{equation}
The probability to find excitation on monomer $n$ is given by the population
$\rho_{nn}(t)$.

In the following, for convenience, we take the number of monomers $N$ to be
even and assume that initially the excitation is located on the two monomers
in the middle of the chain, i.e.\
\begin{equation}
  \label{eq:rho_ini}
\begin{split}
  \rho^{\rm ini}_{N/2,N/2}&=\rho_L\\
\rho^{\rm ini}_{N/2+1,N/2+1}&=\rho_R\\
\rho^{\rm ini}_{N/2,N/2+1}&=\rho_{LR}\\
\rho^{\rm ini}_{N/2+1,N/2}&=\rho_{LR}^{*}\\
\end{split}
\end{equation}
where $\rho_L$ denotes the population on monomer $N/2$ and $\rho_R$ that on
monomer $N/2+1$. The coherence between the two monomers is denoted by $\rho_{LR}$. All other elements are zero.
In matrix notation this can be written as
\begin{equation}
  \label{eq:matrix_ini}
   \rho^{\rm ini}=\left(
\begin{array}{cccc}
\mathbf{0} & \dots&\dots &\mathbf{0} \\
 \vdots &\rho_L& \rho_{LR} & \vdots\\
 \vdots& \rho_{LR}^{*} & \rho_R &\vdots \\
 \mathbf{0}&\dots &\dots & \mathbf{0}
\end{array}
\right)
\end{equation}
We take the initial population on monomer $N/2$ and $N/2+1$ to be equal, i.e.\
$\rho_L=\rho_R=1/2$. Furthermore we write for the coherence $\rho_{LR}=a
\exp({-i\Theta})$, with $a\ge 0$ and $-\pi<\Theta\le \pi$. 
Of particular interest will be the phase factor $\Theta$ between the two monomers. 
For a pure initial state we have $a=1/2$ and the corresponding initial wave function
is given by
\begin{equation}
  \label{eq:init_state}
  \ket{\psi^{\rm ini}}=\frac{1}{\sqrt{2}}\left(\ket{\pi_{\frac{N}{2}}}+ e^{i
      \Theta} \ket{\pi_{\frac{N}{2}+1}} \right).
\end{equation} 
In the following we show how the phase factor $\Theta$ determines the
directionality of excitation transfer.

\section{Numerical calculations}

In this section we will investigate numerically  the time development of the
populations $\rho_{nn}(t)$  by solving \eref{eq:dot_rho_nm}  for different
values of the relative phase $\Theta$ and the dephasing rate $\gamma$.
We take the interaction strength $|V|$ as unit of energy. Consequently we
express time in units of $\hbar/|V|$ and the dephasing rate $\gamma$ in units
of $|V|/\hbar$.

Exemplarily,  in \fref{fig:no_deph} the transfer dynamics is shown for five
values of $\Theta$ and two different $\gamma$ for a chain of $N=60$ monomers.
The coupling $V$ is taken to be positive (changing the sign of $V$ leads to a reversal of the direction of propagation).
Initially the excitation is localized on the sites 30 and 31 with a pure
initial state as given in Eq.~(\ref{eq:init_state}).
In the left column the fully coherent case $\gamma=0$ is shown. In the right
column the case $\gamma=0.3$ is shown. 
In each panel the time dependence of the mean
%In this figure also
%the time dependence of the mean 
\begin{equation}
  \label{eq:M^1}
  M(t)=\sum_n n\rho_{nn}(t)
\end{equation}
is displayed as a black solid line.

Consider first Fig.~\ref{fig:no_deph} c) and h) where $\Theta=0$. Here one clearly sees a
symmetric spreading of the initially localized excitation.
The mean stays constant at its initial location.
 This case is
similar to the situation where the excitation is initially localized only on
one site (see e.g.\ Ref.~\cite{RoEiWo09_058301_}).

Upon increasing the phase difference $\Theta$ to positive values the transfer
to the left becomes enhanced (while the transfer to the right decreases). For
a value $\Theta=+\pi/4$ (see \fref{fig:no_deph} b,g) already a clear
asymmetry can be seen, which becomes stronger when increasing $\Theta$. For
$\Theta=\pi/2$ the transfer to the left reaches its maximum (see
\fref{fig:no_deph}~a,f). 
\begin{widetext} 
\onecolumngrid
\label{sec:numerics}
\begin{figure}[pt]
\psfrag{time}{\small time}
\psfrag{monomer}{$\! \! \! \! \! \! \! \! \! $\small monomer number }
\psfrag{excitation probability}{$\! \! \! \! \! \!\! \! \! \! \! $ excitation probability}
\psfrag{gam0}{\scriptsize $\gamma=0$}
\psfrag{gam0.3}{\scriptsize $\gamma=0.3$}
\psfrag{(a)}{\fontsize{7}{7}\selectfont \bf{(a)}}
\psfrag{(b)}{\fontsize{7}{7}\selectfont \bf{(b)}}
\psfrag{(c)}{\fontsize{7}{7}\selectfont \bf{(c)}}
\psfrag{(d)}{\fontsize{7}{7}\selectfont \bf{(d)}}
\psfrag{(e)}{\fontsize{7}{7}\selectfont \bf{(e)}}
\psfrag{(f)}{\fontsize{7}{7}\selectfont \bf{(f)}}
\psfrag{(g)}{\fontsize{7}{7}\selectfont \bf{(g)}}
\psfrag{(h)}{\fontsize{7}{7}\selectfont \bf{(h)}}
\psfrag{(i)}{\fontsize{7}{7}\selectfont \bf{(i)}}
\psfrag{(j)}{\fontsize{7}{7}\selectfont \bf{(j)}}
\psfrag{Ta}{\scriptsize $\Theta=\frac{\pi}{2}$}
\psfrag{Tb}{\scriptsize $\Theta=\frac{\pi}{4}$}
\psfrag{Tc}{\scriptsize $\Theta=0$}
\psfrag{Td}{\scriptsize $\Theta=-\frac{\pi}{4}$}
\psfrag{Te}{\scriptsize $\Theta=-\frac{\pi}{2}$}
\psfrag{Tf}{\scriptsize $\Theta=\frac{\pi}{2}$}
\psfrag{Tg}{\scriptsize $\Theta=\frac{\pi}{4}$}
\psfrag{Th}{\scriptsize $\Theta=0$}
\psfrag{Ti}{\scriptsize $\Theta=-\frac{\pi}{4}$}
\psfrag{Tj}{\scriptsize $\Theta=-\frac{\pi}{2}$}
\begin{center}
\includegraphics[width=14.9cm]{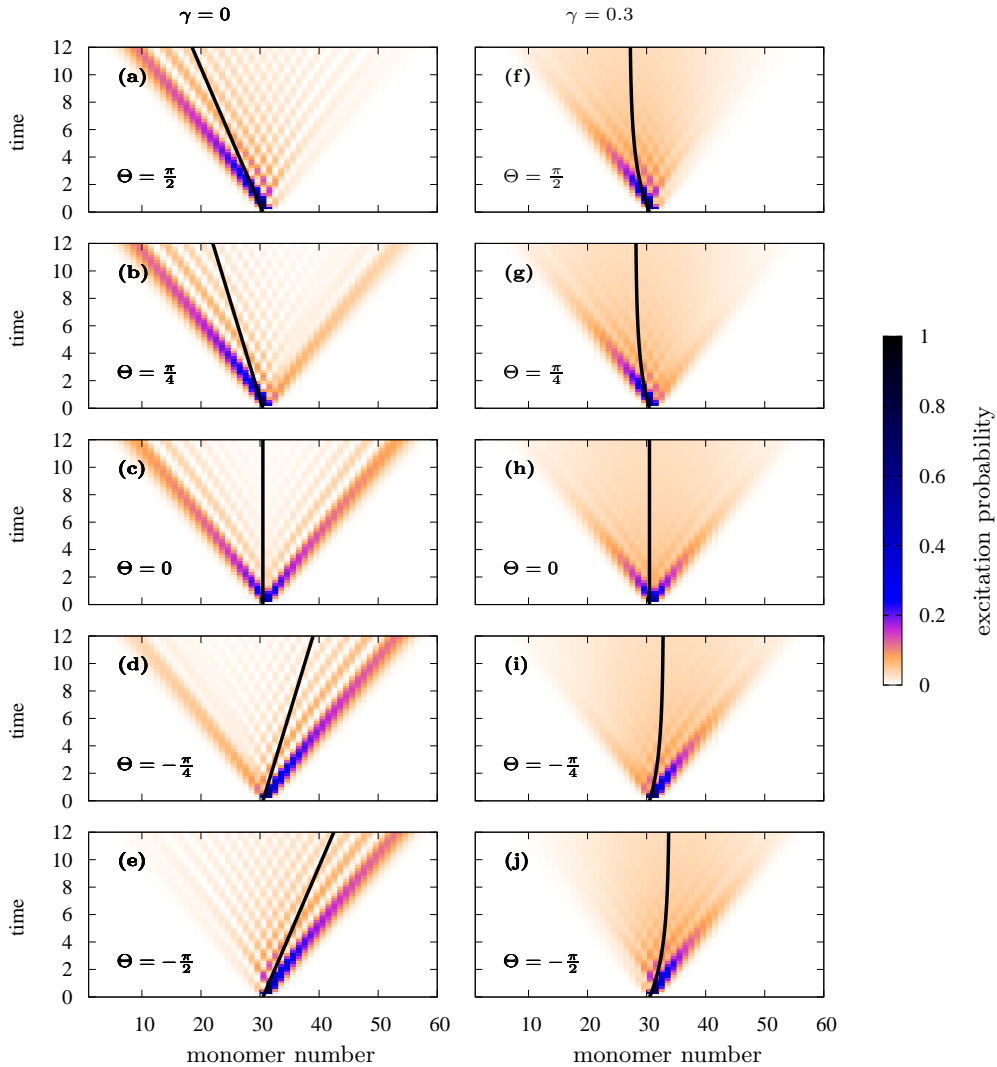}
\end{center}
\caption{\label{fig:no_deph}Probability $\rho_{nn}(t)$ to find excitation on
  monomer $n$ at time $t$ for different values of the phase $\Theta$ and the
  dephasing rate $\gamma$. The black solid line is the time-dependent mean of the
  distribution  $\rho_{nn}(t)$.
Left column $\gamma=0$, right column $\gamma=0.3$.
 Time in units of $\hbar/ V$. The values of $\Theta$ are indicated in the
 figures.}
\end{figure}
\twocolumngrid
\end{widetext}
We have here restricted the discussion to the interval $-\pi/2 \le \Theta \le
\pi/2$. The result for an arbitrary phase $\Theta$ can be
  simply related to the interval $[0,\pi/2]$. For example, for $\Theta>\pi/2$
  we can write $\Theta=\pi/2 +\Theta'$ and we find the same result as for $\Theta=\pi/2 -\Theta'$.
Note that a change of sign of $\Theta$ just reverses the directions. See Fig.~\ref{fig:no_deph}~(d,i) for
$\Theta=-\pi/4$ and Fig.~\ref{fig:no_deph}~(e,j) for $\Theta=-\pi/2$.

While for the case $\gamma=0$ (left column of  \fref{fig:no_deph}) the excitation propagates as a well defined ``wavefront'' over the whole time-range of the propagation, in the case  $\gamma=0.3$ (right column) the exciton dynamics soon becomes similar to a diffusion process.
This behavior is also reflected in the dynamics of the mean.
First we note that one clearly sees  the change of directionality with changing phase $\Theta$.
For the case $\gamma=0$  the mean moves along the chain with constant velocity (characterized by the constant slope of the black line). This velocity  depends on $\Theta$. 
 For a finite dephasing $\gamma=0.3$  and short times the mean also moves with the same
velocity as in the case $\gamma=0$. Soon, however, it slows down and remains
finally quasi constant. We will quantify these observations in the next section.

\begin{figure}[pt]
\psfrag{Theta}{$\Theta$ in $^{\circ}$}
\psfrag{Population}{$ P_L$}%(t\rightarrow \infty)$}
\psfrag{PopR}[c]{\scriptsize $\!\!\!\!\! P_L(t)$}
\psfrag{time}{\scriptsize $t$ in $\hbar/V$}
\includegraphics[width=8.5cm]{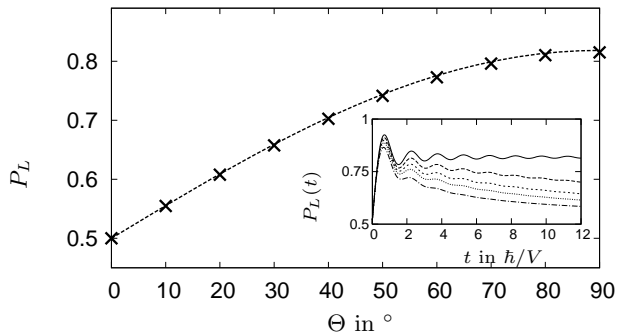}
\caption{\label{fig:Pop_vs_Theta} Asymptotic population on the left side  of
  the chain as a function of the initial phase $\Theta$ for the case of
    vanishing dephasing, i.e.\ $\gamma=0$.
Crosses are the numerical results for $t=12 \hbar/V$, the dashed line is the analytical solution \eref{eq:P_k>0_final}.
The inset shows the population $P_L(t)$ as a function of time for the
  initial phase  $\Theta=\pi/2$ and different values of the dephasing rate
  $\gamma$. From top to bottom the values are 
  $\gamma=0.0$, $\ 0.1$, $\ 0.2$, $\ 0.3$ and  $\ 0.5$. }
\end{figure}
The strong dependence on the phase $\Theta$ can be seen even better by
considering the probability to find excitation on the left or right side of
the chain, i.e.\ by considering the quantities 
\begin{equation}
  \label{eq:P_L}
  P_L(t)\ =\ \sum_{n=1}^{N/2}\rho_{nn}\hspace{0.5cm}{\rm and}\hspace{0.5cm}   P_R(t)\ =\sum_{n=\frac{N}{2}+1}^{N}\rho_{nn}.
\end{equation}
 In Fig.~\ref{fig:Pop_vs_Theta}   the probability
 $P_L(t\!=\!12)$ (i.e.\ to have population on the left side at the final time $t=12$ of \fref{fig:no_deph}), is shown as a function of the initial phase $\Theta$. 
With increasing $\Theta$ initially there is a fast increase in $P_L$ followed
by a quite flat region around the optimum at $\Theta=\pi/2=90^{\circ}$, thus
for small deviations $\Delta \Theta \lesssim 20^{\circ}$ from this optimum
there is still nearly the optimal directionality.  
The inset shows the population $P_L(t)$ as a function of time for the initial phase  $\Theta=\pi/2$ and different
values of the dephasing rate $\gamma$ .

\section{Analytical considerations}
\label{sec:analytics}

\subsection{The fully coherent case}
\label{sec:analytics_coherent}
In this section we give an intuitive explanation for the observed directed
transport.
Since the important contribution to directionality stems from the coherent
motion of the excitation, in the following we will focus on the case $\gamma=0$.
To obtain simple analytical results we restrict to
nearest neighbor interaction and impose periodic boundary conditions (i.e.\  $V_{N,1}=V$).
Then the eigenstates of \eref{eq:Ham} are given by \cite{AmVaGr00__}
\begin{equation}
  \label{eq:ex_state}
  \ket{\phi_k}=\frac{1}{\sqrt{N}}\sum_{n=1}^N\,  e^{i k n}\, \ket{\pi_n}
\end{equation}
with
\begin{equation}
  \label{eq:k}
  k=\frac{2 \pi}{N} j, \hspace{0.5cm}j=-N/2,\dots,N/2-1.
\end{equation}
The corresponding eigenenergies are
\begin{equation}
  \label{eq:E_k}
  E_k=\epsilon+2 V \cos{k}.
\end{equation}
With these results we will now discuss how one can understand directionality
from the initial population of eigenstates $\ket{\phi_k}$.
%\subsubsection{Distribution of exciton-eigenstates in the initial state }
The sign of the wave vector $k$ appearing in the definition of the eigenstates and
eigenenergies can be viewed as the direction of propagation associated with
the stationary state \eref{eq:ex_state}\cite{Fr31_17_}.  For $N\rightarrow
\infty$ the mean velocity associated with the wave vector $k$ is given by
$v_k=\partial_k E(k)= -2V \sin (k)$.  For instance for $k>0$  and positive $V$ the velocity is directed to the left.
Expanding the initial state $\ket{\psi^{\rm
    ini}}$  defined in \eref{eq:init_state} w.r.t.\ the basis states Eq.~(\ref{eq:ex_state}) one finds
\begin{equation}
\ket{\psi^{\rm ini}}=\frac{1}{\sqrt{2 N}}\sum_{k}\, e^{-i k N/2}\big( 1+ e^{i
  (\Theta -k)}\big)\, \ket{\phi_k}.
\end{equation}
 The probability to find a certain $k$ within the initial state $\ket{\psi^{\rm
    ini}}$  (defined in \eref{eq:init_state}) is given by
 \begin{equation}
    \label{eq:prob_k}
   \begin{split}
    P_k=|\braket{\phi_k}{\psi^{\rm ini}}|^2
    =& \frac{1}{N}\big(1+\cos(k-\Theta)).
    \end{split}
  \end{equation}
From this equation one sees that for $\Theta=0$ the probability to find
positive and negative $k$-values is equal.
For negative values of $\Theta$ the maximum of the function $P_k$ is shifted to positive
$k$, i.e.\ velocities directed to the right for $V>0$.
This has consequences also for the initial velocity
\begin{equation}
\begin{split}
\label{eq:v_ini_aus_k}
v^{\rm ini}=&\sum_k P_k v_k 
=   - V  \sin (\Theta),
\end{split}
\end{equation}
which shows a sinusoidal dependence on the phase $\Theta$. We will come back to this result in the next section.

Consider now the probability to find
positive $k$-values in the initial state
\begin{equation}
  \label{eq:P_k>0}
 \begin{split}
  P_{k>0}=&\sum_{j=1}^{N/2-1} P_{\frac{2 \pi}{N} j}
=\frac{1}{2}- \frac{1}{N}+\frac{\cot\big(\frac{\pi}{N}\big)}{N}\, \sin(\Theta).
\end{split}
\end{equation}
 In the limit $N\rightarrow \infty$ \eref{eq:P_k>0}  simplifies to
\begin{equation}
  \label{eq:P_k>0_final}
  P_{k>0}\approx \frac{1}{2}+\frac{\sin(\Theta)}{\pi}
\end{equation}
This analytical function is shown in \fref{fig:Pop_vs_Theta} as a dashed line.
It is in very good agreement with the asymptotic populations (crosses in
\fref{fig:Pop_vs_Theta}) found numerically  for
the case $\gamma=0$. 
Thus $P_{k>0}$ is just the population on the left side of the chain at large times, i.e.\ $P_{L}(t\rightarrow\infty)=P_{k>0}$.
For a finite chain this result holds only for times before
  the first reflection at the end of the chain occurs.  Note that the time has
  to be large enough so that the initial
  oscillations shown in the inset of \fref{fig:Pop_vs_Theta} do not play a role
  anymore.

\subsection{The first moment of exciton motion}
As seen in Section \ref{sec:numerics} the mean position
  of the excitation $M(t)$, defined in Eq.~(\ref{eq:M^1}),
is a useful measure to quantify directionality.
For our initial condition the excitation is localized on monomer $N/2$ and
$N/2+1$ with equal probability. 
Thus, $M(t)>(N+1)/2$  means that the excitation is mainly
localized on the right side of the chain, for $M(t)<(N+1)/2$ it can mainly  be
found on the left.

Following the treatment of Reineker \cite{ReKr82__} we can find a differential
equation for $M(t)$.
In the following, for simplicity, we assume  nearest neighbor interactions only
and write $V_{nm}=V (\delta_{m,n+1}+\delta_{m,n-1})$ with equal interaction
between all neighboring monomers.
Furthermore we consider an infinite chain.
Then, after some algebra, one finds
\begin{equation}
  \label{eq:dot_M1}
  \frac{d}{dt} M(t) = V \cdot \phi(t)
\end{equation}
with 
\begin{equation}
  \label{eq:phi_m}
  \phi(t)=  i\, \sum_n(\rho_{n+1,n}-\rho_{n,n+1})
\end{equation}
Differentiating $\phi(t)$ w.r.t.\ time and noting that $\sum_n
\rho_{nn}=\sum_n \rho_{n+1,n+1}$ and $\sum_n
\rho_{n+2,n}=\sum_n \rho_{n+1,n-1}$, from Eqs.~(\ref{eq:dot_rho_nm}) and~(\ref{eq:phi_m}) we get
\begin{equation}
  \label{eq:dot_phi_1}
 \frac{d}{dt} {\phi}(t)= -\gamma \phi(t).
\end{equation}
This equation can be easily integrated to give
 \begin{equation}
   \label{eq:lsg_phi_1}
   \phi(t)=\phi(0) e^{-\gamma t}
 \end{equation}
with $\phi(0)$ determined by the initial density matrix $\rho^{\rm ini}$.
Note that for our 
special initial condition Eq.~(\ref{eq:rho_ini}) with $\rho_{LR}=a
\exp({-i\Theta})$ one has
 $ \phi(0)=2\, {\rm Im} \rho_{LR}= -2\, a\, \sin \Theta$. 
Inserting the result \eref{eq:lsg_phi_1}  into Eq.~(\ref{eq:dot_M1}) gives 
\begin{equation}
  \label{eq:lsg_M_1(t)}
  M(t)=M(0)+\left\{
\begin{array}{l}
V  \phi(0) \cdot  t \hspace{1.7cm} {\rm for}\hspace{0.2 cm} \gamma=0\\
\\
\frac{V}{\gamma} \phi(0) (1- e^{-\gamma t}) \hspace{0.55 cm} {\rm for}\hspace{0.2 cm}\gamma \ne 0
\end{array}
\right.
\end{equation}
From  Eq.~(\ref{eq:lsg_M_1(t)}) one sees that the initial density matrix $\rho^{\rm ini}$ (entering via
$\phi(0)$) plays a crucial role in the time evolution of $M(t)$. 
If $V \phi(0)>0$ the  mean  $M(t)$ will move to larger $n$.
Note that a change of the sign of $V$ also changes the direction.

Before considering the initial condition Eq.~(\ref{eq:rho_ini}) in more
detail, let us first discuss the dependence on the dephasing rate $\gamma$.
From Eq.~(\ref{eq:lsg_M_1(t)}) one sees that for $\gamma=0$ the mean $M(t)$
moves with a constant velocity given by $|V \phi(0)|$, which  for a pure state (where $\phi(0)=-\sin\Theta$) is in agreement with Eq.~(\ref{eq:v_ini_aus_k}) obtained by considering the distribution of $k$-values in the initial state.
For $\gamma\ne 0$ this velocity becomes time-dependent.
As long as $t\ll 1/\gamma$ one can expand Eq.~(\ref{eq:lsg_M_1(t)}) to find 
$M(t) \approx M(0)+ V \phi(0) t +\dots$.
Thus initially the mean moves with the same velocity as without dephasing.
For $t \gtrsim 1/\gamma$  the directionality of the transfer is mostly lost and
the mean $M(t)$ slowly approaches its long time limit
\begin{equation}
  \label{eq:long_time}
  M(t \gg 1/\gamma)=M(0)+\frac{V}{\gamma} \phi(0)
\end{equation}

After this discussion of some general properties of $M(t)$ we go back to the
special initial condition Eq.~(\ref{eq:rho_ini}) with $\rho_{LR}=a
\exp({-i\Theta})$.
Then, as already noted above,
one has $ \phi(0)=-2\, a\, \sin \Theta$. 
For given $V$ and $a$ the direction of transport is solely determined by the phase $\Theta$.
 We find that for $\Theta<0$ the transport is
directed to the right  (for positive $a$ and $V$) with maximum at $-\pi/2$  while for $0<\Theta<\pi$ it is directed to the left.
For $\Theta=0$ the situation is symmetric and the mean remains $M(0)$.
Note that $a$ scales the ``strength'' of directionality.
For a fully mixed state ($a=0$) there is no directionality; for a pure initial
state ($a=1/2$) the directionality is maximal.

\section{Summary and Conclusions}
\label{sec:conclusion}
In the present work we have investigated how the direction of excitation
transfer along a one-dimensional aggregate depends on the relative phase
between two initially excited monomers.
It was found that in the case, when there is no interaction with the
environment, for a phase difference of $\pi/2$ more than 80\% of the excitation
propagates into one direction. The direction depends on the sign of the
phase.
For smaller phase differences the directionality decreases. The same holds, if the initial state is not pure. 
As expected interaction with an environment, which destroys phase relations
between excitations on different monomers, leads to a strong decrease of the
directionality. 
In this case, the directionality of the propagating wave-packet is only conserved on timescales of the order of the
dephasing time.
This dependence on the initial phase is also reflected in the time-dependence of  the mean of the excitation. 
For this quantity exact analytical results have been derived, showing in
particular that
initially the  velocity of the mean is given by the product of the transfer-interaction between the monomers and the sine of the phase-difference in the initial state. In the   fully coherent case this velocity remains constant;  with dephasing the velocity  decreases exponentially on the timescale of the
dephasing time. 
While fluctuations of the environment can often be helpful to enhance the
transport efficiency of excitonic energy \cite{PlHu08_113019_,ReMoKa09_033003_}, in the present
case the coherence is essential for the directed transport to persist.
In this context it is also appropriate to briefly discuss the influence
of internal vibrations of the monomers as they are present e.g.\ in aggregates
of organic dyes \cite{EiBr06_376_}.
We have performed preliminary calculations to obtain a feeling how the
directionality is affected by these vibrations. For a single undamped
vibrational mode per monomer, we used the formalism of
Ref.~\cite{RoScEi09_044909_} which is based on the ``coherent exciton
scattering'' approximation which allows to treat chains up to hundred monomers. 
%For chains with a length of ten monomers we have checked with a more
 %accurate method developed in Ref.\cite{RodenPTCDA}.
 We found that the inclusion of a single vibrational mode per monomer barely
 alters the  degree of directionality compared to the purely electronic case of Section~\ref{sec:analytics_coherent}. 
However, similar to the findings in Ref.~\cite{RoScEi09_044909_} the velocity
of propagation is slowed down. To assess the influence  of many (damped) vibrational modes we used the non-Markovian Quantum State Diffusion (NMQSD) approach of Ref.~\cite{RoEiWo09_058301_}, which allows to treat continuous spectral densities. 
For a Markovian environment the NMQSD approach can be shown to be equivalent
to the master equation (\ref{eq:dot_rho_nm}). Considering environments that are no longer Markovian (e.g.\ weakly damped vibrational modes) the resulting transfer has a complicated behavior showing  both dephasing as well as the influence of individual vibrational modes, which requires further investigations.
However, our investigations indicate that the inclusion of vibrational modes
does not substantially alter the results found in the present article.

\begin{figure}[tp]
\psfrag{Vnm}{\footnotesize $V_{nn+1}$}
\psfrag{time}{\tiny time [arb.u.]}
\includegraphics[width=8.9cm]{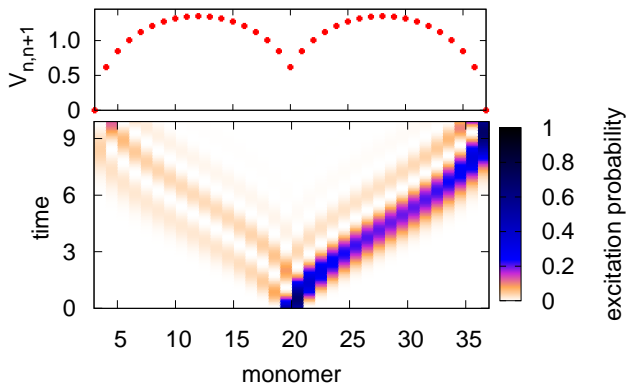}
\caption{\label{fig:optimal}Engineered interaction between the
  monomers. \\ a) Strength of the interaction between monomer $n$ and $n+1$.  b)
  Corresponding transfer for an initial state with $\Theta=-\pi/2$.  }
\end{figure}
Note that for initial states that consist of coherent superpositions of a
large number of monomers a better directionality is possible.
For example, we found that when the excitation is delocalised over four monomers the
directionality can already exceed 90\%.
Clearly, when the number of monomers that initially are   excited coherently
becomes even larger a nearly perfect directionality is possible, e.g. for a
Gaussian distribution of $k$ values around a certain mean wave vector $k_0$
(this results in a Gaussian wave packet in $n$-space).
The present study shows, that even for the smallest possible superposition (involving excitation of only two monomers)
pronounced directionality is possible. 
In this regard it is noteworthy that the directed transfer also works for
arrangements where the interaction between neighboring monomers is no longer
equal. In nature this is the case e.g.\ in the LH2 complex of purple bacteria. In artificial systems one might design the couplings (e.g.\ by changing the distance or orientations between the monomers) such that the propagating wave packet has certain properties like a minimal spread over the monomers or the refocusing at a certain monomer. That  such tailoring of the aggregate properties can be combined with the directionality induced by the initial phase is  exemplarily  shown in Fig.~\ref{fig:optimal} for nearest
neighbor interactions as given in panel (a)\footnote{The interactions on each
  half are chosen in a way as discussed e.g.\ in Ref~\cite{ChDaEk04_187902_}.},
chosen such that the initial excitation is focused at the end points
  of the chain.
This
 might stimulate  new ideas how to employ the excitonic phase in molecular devices. 
The directionality imposed by the relative phase in coherent superpositions
of localized states might even play a role in the energy transfer in natural light harvesting
systems\cite{NoRa96_203_}, when partly delocalized excitations ``hop'' from one complex to the next
and the exciton remains delocalized over a small number of monomers. 

\begin{acknowledgements}
Discussions with J.~Roden (who also provided the  computer programs to
investigate effects of vibrations), G.~Ritschel, S.~M\"obius, C.~Hofmann, Mr.\ Saalmann and Dr.~Croy  are gratefully acknowledged. 
 
\end{acknowledgements}

%\bibliography{PhaseDir.bib}
%%%%%%%%%%%%%%%%%%%%%%%%%%%%%%%%%55
%
%%%%%%%%%%%%%%%%%%%%%%%%%%%%%%%%%%%%%%%

\end{document}